\documentclass[preprint,showpacs,preprintnumbers,amsmath,amssymb]{revtex4}
\usepackage{amssymb}
\usepackage{graphicx}
\usepackage{dcolumn}
\usepackage{bm}

\begin{document}

\title{Magnitude of Magnetic Field Dependence of a Possible Selective Spin
Filter in ZnSe/Zn$_{1-x}$Mn$_{x}$Se Multilayer Heterostructure}
\author{Zhen-Gang Zhu$^{1\ast }$ and Gang Su$^{2}$}
\affiliation{$^{1}$Center for Advanced Study, Tsinghua University, Beijing 100084, China\\
$^{2}$Department of Physics, The Graduate School of the Chinese Academy of
Sciences, P. O. Box 3908, Beijing 100039, China}

\begin{abstract}
Spin-polarized transport through a band-gap-matched ZnSe/Zn$_{1-x}$Mn$_{x}$%
Se/ZnSe/Zn$_{1-x}$Mn$_{x}$Se/ZnSe multilayer structure is investigated. The
resonant transport is shown to occur at different energies for different
spins owing to the split of spin subbands in the paramagnetic layers. It is
found that the polarization of current density can be reversed in a certain
range of magnetic field, with the peak of polarization moving towards a
stronger magnetic field for increasing the width of central ZnSe layer while
shifting towards an opposite direction for increasing the width of
paramagnetic layer. The reversal is limited in a small-size system. A strong
suppression of the spin up component of the current density is present at
high magnetic field. It is expected that such a reversal of the polarization
could act as a possible mechanism for a selective spin filter device.
\end{abstract}

\maketitle

It is indispensable to realize the effective spin-polarized
electrons injection (spin injection) into semiconductor for
spin-related semiconductor devices, such as spin transistors
\cite{datta}, etc. Two groups have been able to demonstrate the
efficient spin injection into GaAs using semimagnetic $BeMnZnSe$
\cite{fiederling} and ferromagnetic $GaMnAs$ epilayers \cite{ohno},
respectively. Physics governing spin injection and detection into
semiconductor is now being understood \cite{rashba,schmidt1,fert}.
More recently, new attempts to realize the devices where the spin
character of the injected and detected electrons could be voltage
selected \cite{schmidt2}, have been made. A magnetic resonant
tunneling diode (RTD) is considered, in which the semimagnetic
Zn$_{1-x}$Mn$_{x}$Se is used as the spin-splitted well.
Bias-dependent current polarization or spin filter can be expected,
and the results demonstrate the possibility of devices based on
tunneling through spin resolved energy levels. Theoretically, the
spin-dependent tunneling through a similar structure is investigated
by Sugakov \textit{et al.} \cite{sugakov}, and Egues \textit{et al}.
\cite{egues1}. In this paper, we shall explore a magnetic field
tunable structure, ZnSe/Zn$_{1-x}$Mn$_{x}$
Se/ZnSe/Zn$_{1-x}$Mn$_{x}$Se/ZnSe. The Mn concentration in the
paramagnetic layer (PL) is chosen so that the offsets of conduction
and valence bands are nearly zero in the absence of an applied
magnetic field, and the alternation of ZnSe and Zn$_{1-x}$Mn$_{x}$Se
may be used to build up the spin superlattice. The results show an
interesting phenomenon that reversal polarization of current density
can be induced by tuning the magnitude, not its direction, of
magnetic field because of resonant tunneling effect. In other words,
spin injection can be selected by the magnitude of magnetic field.

Mn- or Fe-based spin superlattices were proposed by von Ortenberg
\cite{ortenberg}, and realized by Chou \textit{et al}. \cite{chou}
and Dai \textit{et al}. \cite{dai}. Time-resolved photoluminescence
spectroscopy was used to investigate exciton lifetime and spin
relaxation in magnetic semiconductor spin superlattices
\cite{smyth}. Egues \cite{egues2} investigated spin filtering in a
ZnSe/Zn$_{1-x}$Mn$_{x}$Se heterojunction with a single paramagnetic
layer (SPL), and observed a strong suppression of the spin up
component of the current density for increasing magnetic field. Guo
\textit{et al.} \cite{guo} investigated the bias-dependent spin
transport in a SPL structure and spin filtering in
ZnSe/Zn$_{1-x}$Mn$_{x}$Se with double paramagnetic layers (DPL).
Theoretical investigations, as above mentioned, have not given a
complete description about the size-dependence of spin transport,
while the spin-dependent transport in these heterostructures is
quite sensitive to size. So, it is our purpose to focus on this
issue, and a DPL structure will be considered below.

Let us consider the conduction electron transport through a magnetic
field tunable heterojunction such as
ZnSe/Zn$_{1-x}$Mn$_{x}$Se/ZnSe/Zn$_{1-x}$Mn$_{x}$ Se/ZnSe with DPL,
in which \textit{sp-d} exchange interaction gives rise to a
spin-dependent potential \cite{furdyna,nalwa}
$V_{\sigma_{z}}(z)=-xN_{0}\alpha \sigma _{z}\left\langle
S_{z}\right\rangle [\Theta (z+W_{L})\Theta (-z)+\Theta (z-b)\Theta
(b+W_{R}-z)]$ in the Hamiltonian of the system. Here $N_{0}\alpha $
is the electron \textit{s-d} exchange constant, $x$ is the Mn
concentration, $\sigma _{z}$ are the electron spin components $\pm
1/2$ (or $\uparrow $, $\downarrow $) along the field, $\left\langle
S_{z}\right\rangle $ is the thermal average of the z-th component of
a Mn$^{2+}$ spin (a 5/2 Brillouin function), $\Theta (z)$ is the
Heaviside function, $W_{L(R)}$ is the width of left (right)
Zn$_{1-x}$Mn$_{x}$Se layer, and $b$ is the width of the central ZnSe
layer. The transmission coefficient (TC) of the single
Zn$_{1-x}$Mn$_{x}$Se layer structure can be obtained by using the
transfer matrix method \cite{ferry,davies}. The result is
\begin{equation}
T_{tot}^{\uparrow (\downarrow )}(E)=\frac{T_{L}^{\uparrow (\downarrow
)}T_{R}^{\uparrow (\downarrow )}}{(1-\sqrt{R_{L}^{\uparrow (\downarrow
)}R_{R}^{\uparrow (\downarrow )}})^{2}+4\sqrt{R_{L}^{\uparrow (\downarrow
)}R_{R}^{\uparrow (\downarrow )}}\cos ^{2}\phi _{\uparrow (\downarrow )}},
\label{ttot}
\end{equation}%
where $\phi _{\uparrow (\downarrow )}=\theta ^{\uparrow (\downarrow
)}/2-kb$ with $\theta ^{\uparrow (\downarrow )}$ the phase of the
$11$ element of the transfer matrix of the SPL structure, and $k$
the wavevector in the ZnSe layer (we neglect the spin-dependent of
the wavevector in layer ZnSe, for the split of the spin up and down
induced by a magnetic field is small), $T_{L}^{\uparrow (\downarrow
)}$ is the TC for a SPL structure and $L$ denotes the left PL in the
DPL structure, $R_{L}^{\uparrow (\downarrow )}$ is the reflection
coefficient. In the following, some parameters are assumed: the
effective mass of electron $m_{e}^{\ast }=0.16m_{e}$ with $m_{e} $
the mass of bare electron, $W_{L}=W_{R}=W$, and an effective Mn
concentration $x_{eff}=x(1-x)^{12}$ to account for the
antiferromagnetic clustering effects \cite{egues2}.

The degeneracy of spin subbands is removed because of the
\textit{sp-d} exchange interaction in a PL, and a spin-dependent
potential is induced. Up(down)-spin electrons see a barrier (well)
in a PL. For a DPL structure, up(down)-spin electrons see a
double-barrier (-well) structure (DB(W)S). The system under interest
is a combination of DBS and DWS for different spin orientations
simultaneously. Different phases $\theta ^{\uparrow (\downarrow )}$
for a SPL structure give rise to different phases $\phi _{\uparrow
(\downarrow )}$ for a DPL structure. The resonant transport can
occur at $\phi _{\uparrow (\downarrow )}=(2n+1)\pi /2$. This
condition determines a splitting resonant energies
$E_{n}^{\uparrow}$ and $E_{n}^{\downarrow }$ for up-spin and
down-spin electrons, respectively.
\begin{figure}[tbph]
\centering \includegraphics[width =15 cm, height=10 cm]{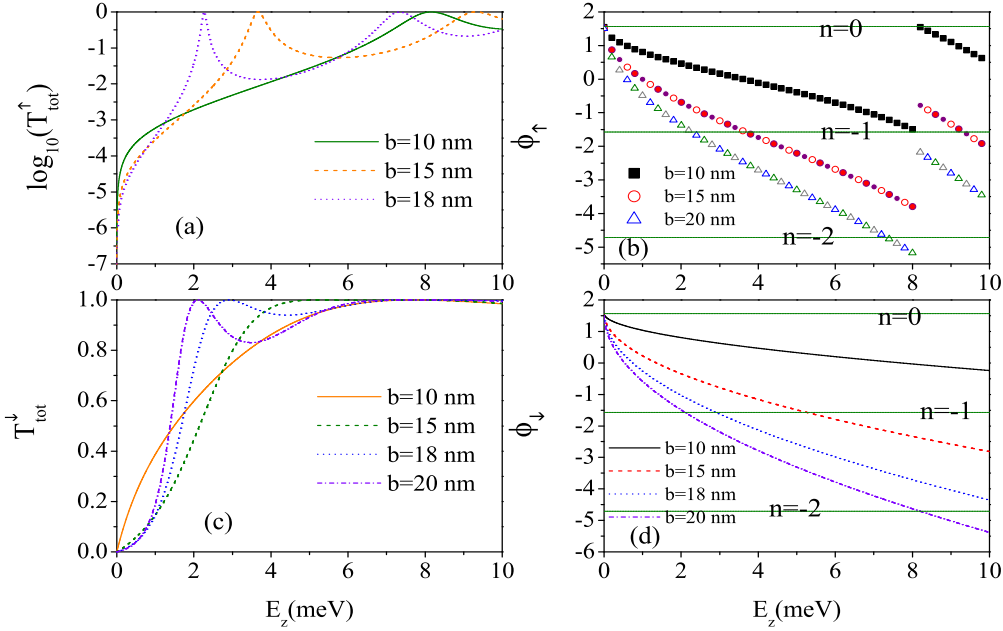}
\caption{(color on line) The transmission coefficient
$T_{tot}^{\uparrow (\downarrow )}$ and the phase $\phi _{\uparrow
(\downarrow )}$ as functions of electron energy $E_{z}$. (a)
$\log_{10}(T_{tot}^{\uparrow })$; (b) $\phi _{\uparrow }$; (c)
$T_{tot}^{\downarrow }$; (d) $\phi _{\downarrow }$. The parameters
are taken as \cite{furdyna,nalwa} $W=100$ \AA , $B=1$ T, $S=5/2$,
$\Theta =1.32$ (Curie-Weiss temperature), $N_{0}\alpha =0.26$ and
$x=0.06$. } \label{fig1}
\end{figure}

For incident energies $E_{z}<x\left\vert \left\langle
S_{z}\right\rangle \right\vert N_{0}\alpha /2$, the wave of up-spin
electrons is evanescent in the PL. However, the TC can be
considerable for a DBS at the resonant tunneling case, which can
even approach to unity when the system is symmetric. This feature is
obvious, as shown in Fig. 1(a), and the lines of $n=0$, $n=-1$ and
$n=-2$ are displayed in Fig. 1(b). The crossing points of $\phi
_{\uparrow }$ and these lines will determine the resonant tunneling
energies or the quasi-bound states in the central ZnSe well. With
increasing $b$, the position of the quasi-bound states goes down
(see $E_{-1}^{\uparrow }$). This tendency gives rise to a shift of
the resonant peaks of TCs to lower energies with increasing $b$, as
shown in Fig. 1(a). For instance, when $b=100$ \AA , $150$ \AA ,
$180$ \AA , then $E_{0}^{\uparrow }=8.155$ meV, $3.663$ meV, $2.257$
meV, respectively. Recall that this feature is also present in the
conventional semiconductor RTD \cite{ferry,tsuchiya}.

The energy-dependence of $T_{tot}^{\downarrow }$ and $\phi
_{\downarrow }$ is depicted in Figs. 1(c) and (d), respectively. The
TCs increase with increasing $E_{z}$ and approach to unity when the
conditions $\phi _{\downarrow }=(2n+1)\pi /2$ or
$T_{L(R)}^{\downarrow }=1$ and $b=W_{L(R)}=W$ are satisfied (as in
the case $b=100$ \AA ), while the resonant peaks move to lower
energies: as $b=150$ \AA , $180$ \AA , $200$ \AA ,
$E_{-1}^{\downarrow }=5.257$ meV, $2.892$ meV, and $2.099$ meV,
respectively. Because of the split of resonant energies
$E_{n}^{\uparrow }$ and $E_{n}^{\downarrow }$, the TC of up-spin
electrons at the resonant energy may be larger than that of
down-spin electrons at the same energy (which may be the
off-resonant energies for down-spin electrons). This character will
lead to a split of the current density for different spin
orientations, and may be the origin of the reversal of polarization
(see below).
\begin{figure}[tbph]
\centering \includegraphics[width =15 cm, height=10 cm]{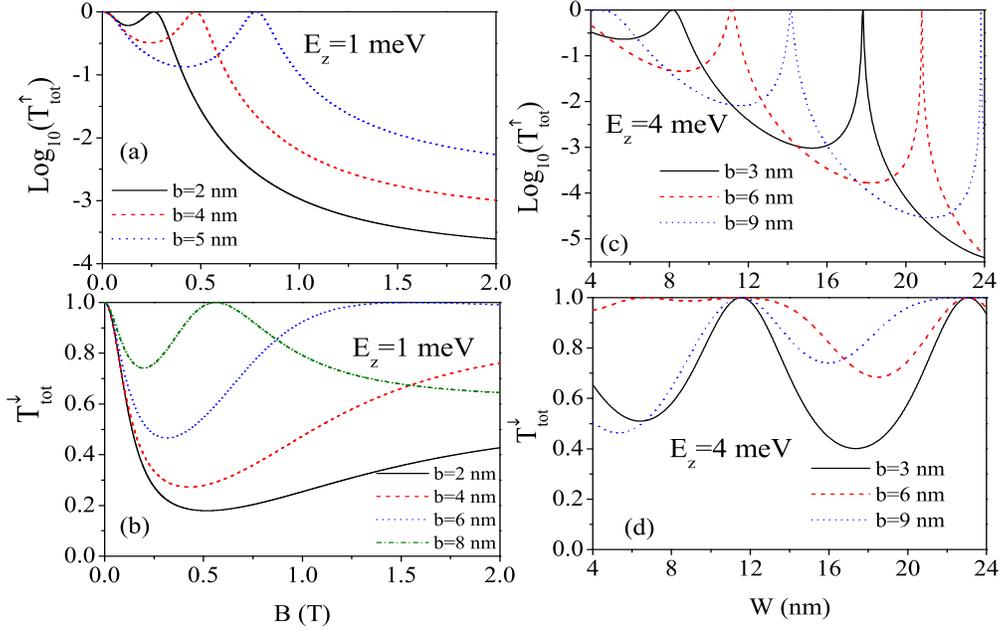}
\caption{(color on line) The magnetic field dependence of
$\log_{10}(T_{tot}^{\uparrow })$ (a) and $T_{tot}^{\downarrow }$
(b), where $W=100$ \AA , and $E_{z}=1$ meV. The transmission
coefficient $\log_{10}(T_{tot}^{\uparrow })$ (c) and
$T_{tot}^{\downarrow }$ (d) versus the width of the paramagnetic
layer $W$ for different $b$, where $B=1$ T, and $E_{z}=4$ meV. The
other parameters are the same as those in Fig. 1.} \label{fig2}
\end{figure}

Increasing the magnetic field will lead to a higher barrier for
up-spin electrons while to a deeper well for down-spin electrons,
giving rise to the probability that the up-spin electrons penetrate
into the barrier will be lowered. As a result, the up-spin electrons
prefer staying in the ZnSe layer, while down-spin electrons prefer
staying in the semimagnetic semiconductor layer, coined as spin
superlattice \cite{ortenberg,chou,dai}. This manifests itself as the
decreasing of TC for up spin except the resonant peaks and may raise
the positions of the quasi-bound states as well as the phase
$\phi_{\uparrow }$ in the central ZnSe layer. For wider central ZnSe
layer, stronger magnetic field is needed to raise the quasi-bound
states to match the incident energy and generate a resonance. So the
resonant peaks of $T_{tot}^{\uparrow }$ shift to a larger $B$ with
increasing $b$, as shown in Fig. 2(a). The variation of
$T_{tot}^{\downarrow }$ with $B$ is depicted in Fig. 2(b). The
$b$-dependence of the resonant peaks for down-spin electrons seems
to be opposite to that for up-spin electrons, i.e. the peaks shift
towards lower $B$. It is observed that the phase $\phi
_{\downarrow}$ decreases with increasing the magnetic field.

The $W$ dependence of $T_{tot}^{\uparrow ,\downarrow }$ is shown in Figs.
2(c) and (d). Apart from a few sharp resonances, $T_{tot}^{\uparrow }$ is
overall decreasing with increasing $W$ because of the evanescent wave in the
barrier. For down-spin electrons, $T_{tot}^{\downarrow }$ is oscillating
with increasing $W$, and the crossing points where the total TCs are unity,
are observed at particular $W$s for different $b$. This latter property
occurs because the TCs of SPL structure are unity (i.e. $T_{L}^{\downarrow
}=T_{R}^{\downarrow }=1$), implying that the SPL is completely transparent
for down-spin electrons, and the electrons travel through it without any
reflections.

Now let us investigate the polarized current density through the DPL
structure. It is necessary to sum the contribution of discrete
Landau levels $n$ which are filled to Fermi energy. The
spin-dependent current density $J_{\uparrow }$ $(J_{\downarrow })$
can be calculated in the manner similar to that in Ref.
\cite{egues2}. A small-bias limit, i.e. $eV=E_{f}\approx 5$ meV, is
assumed for numerical calculation, and let $J_{0}\equiv e^{2}/4\pi
^{2}\hbar ^{2}c$. To evaluate the spin-polarized effect on the
current density, it is useful to get the spin polarization of the
transmitted beam which is defined as
\begin{equation}
P=\frac{J_{\uparrow }(B)-J_{\downarrow }(B)}{J_{\uparrow }(B)+J_{\downarrow
}(B)}.  \label{po}
\end{equation}

\begin{figure}[tbph]
\centering \includegraphics[width =15 cm, height=10 cm]{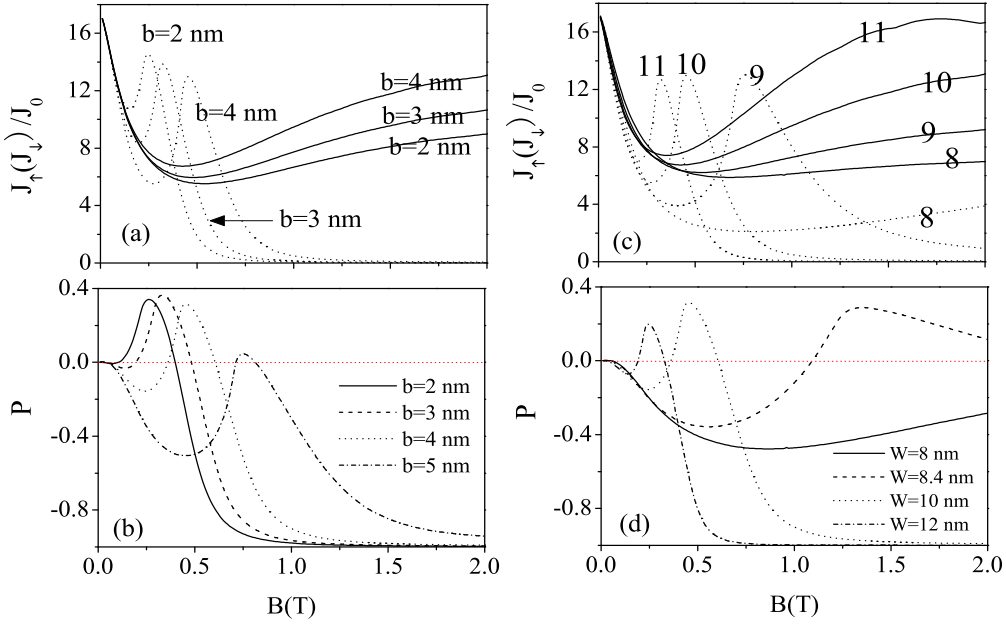}
\caption{(color on line) Current density $J_{\uparrow }$
$(J_{\downarrow })$ and the polarization $P$ versus the magnetic
field for different $b$ in (a) and (b) and for different $W$ in (c)
and (d), dot (solid) lines for $J_{\uparrow }$ $(J_{\downarrow })$
in (a) and (c). In (a), the digits sitting on the curves in (a) are
magnitudes of $b$ correspond to $b=2,3,4$ nm, respectively. And
$W=100$ \AA\ in (a) and (b). In (c), the digits correspond to
$W=8,9,10,11$ nm, respectively. $b=40$ \AA\ in (c) and (d).
$E_{f}=5$ meV for all the four figures and the other parameters are
the same as those in Fig. 1.} \label{fig3}
\end{figure}
The magnetic field dependence of the current density $J_{\uparrow }$
$(J_{\downarrow })$ and the current polarization $P$ are shown in
Fig. 3. The variation of the current density with the magnetic field
under consideration is not very similar to that in Ref. \cite{guo},
but it also depends closely on features of $T_{tot}^{\uparrow
(\downarrow )}$. $J_{\uparrow }$ (dotted lines in Fig. 3(a) and (c))
first decreases, then goes to a maximum, and then decreases almost
to vanishing with increasing $B$. The quasi-bound resonances
manifest themselves in $J_{\uparrow }$ by resonant peaks which vary
with $b$. The resonant peaks corresponding to larger $b$ appear at
larger $B$. However, for a very large $b$ (e.g. $b=500$ \AA\
\cite{guo} or $1000$ \AA\ \cite{egues2}), the current density for
spin up decreases exponentially with increasing the magnetic field
and the resonant peaks are almost vanished. $J_{\downarrow }$ (solid
lines in Fig. 3(a) and (c)) is not strongly suppressed by $B$
because the wave functions of down-spin electrons are traveling
waves. The suppression of up-spin component leads almost to a
perfect spin filter effect at a stronger magnetic field \cite{guo},
i.e. only down-spin electrons transmit and $P$ is almost equal to
$-1$ (Fig. 3(b)).

The split of $J_{\uparrow }$ and $J_{\downarrow }$ induced by the
split of the phases of different spin components leads to the
polarization of the transmitted current density. It makes
$J_{\uparrow }$ larger than $J_{\downarrow }$ at certain range of
$b$, and $P$ will be positive, as shown in Fig. 3(b). It can be seen
that the polarization $P$ is reversed from negative to positive for
increasing $B$, and approaches to its maximum, then decreases to
$-1$. This reversal of $P$ exists in a distinct wide range of $B $
for appropriate $b$. $P$ tends to $-1$ for a stronger magnetic
field, suggesting a perfect spin filter effect. The peaks of $P$
move to larger $B$ for increasing $b$ corresponding to the shift of
the resonant peaks of TC with $B$ in Fig. 2(a). The similar $B$
dependence of $J_{\uparrow }$ $(J_{\downarrow })$ and $P$ are shown
in Fig. 3(c) and (d) at different $W$. It can be found that the
resonant peaks move to weaker magnetic field for increasing $W$.
This tendency is just opposite to the shift of peaks with increasing
$b$ in Fig. 3(a) and (b). This is owing to the opposite $W$ and $b $
dependence of the phase $\phi $. The reversal of the polarization
disappear when the width of PL is small and very large.

The positive $P$ exists in a certain size of the system. To
substantiate this point, we show a a contour plot of a 3D graph in
which the $z$ axis is the largest $P$ in a range $B\in \lbrack
0,2T]$ for a fixed $b$ and $W$ in Fig. 4. A positive maximum $P$\
means that there is the reversal of $P$\ in the range $B\in
\lbrack 0,2T] $ which is the bright region in Fig. 4. On the
contrary, the dark region means there is no the reversal of the
polarization of the current. It is interesting to note that the
reversal is limited to a finite region where the layers should be
thin, and thicker layers may suppress this effect. It may be
caused by the splitting of the resonant transport for different
spin orientations.
\begin{figure}[tbph]
\centering \includegraphics[width =15 cm, height=10 cm]{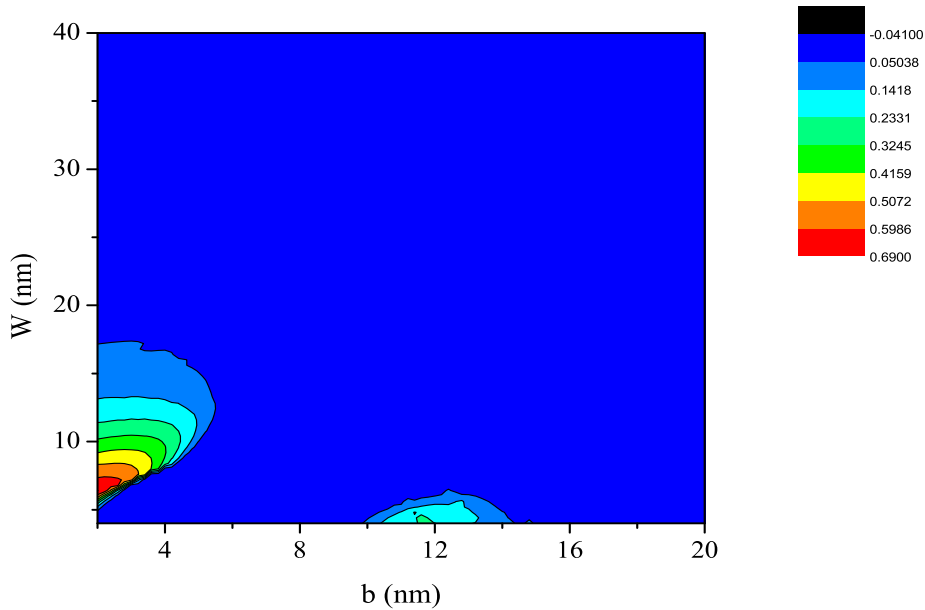}
\caption{(color on line) A contour plot of the maximum $P$ in
$[0,2$T$]$ versus $b$ and $W$, where $E_{f}=5$ meV, and the others
are the same as those in Fig. 1.} \label{fig4}
\end{figure}

In summary, the spin-polarized transport through a band-gap-matched
ZnSe/Zn$_{1-x}$Mn$_{x}$Se/ZnSe/Zn$_{1-x}$Mn$_{x}$Se/ZnSe multilayer
structure is investigated. In an external magnetic field, the
paramagnetic layers serve as barriers for up-spin electrons and
wells for down-spin electrons simultaneously. The system under
interest is then the combination of a DBS for up-spin electrons and
a DWS for down-spin electrons. The resonant transport will occur at
different energies for electrons with different spins because of the
split of the energy levels in the layers. The phases $\phi
_{\uparrow ,\downarrow }$ also split, and $\phi _{\uparrow
(\downarrow )}=(2n+1)\pi /2$ determines the resonant energies
manifested itself in the transmission coefficients for up-spin and
down-spin electrons. The consequent result is that the polarization
of the current density can be reversed in a certain range of
magnetic fields. It should be pointed out that the reversal results
from the splitting of the resonant transport through DBS and DWS.
This reversal mechanism is different from that presented in Ref.
\cite{schmidt2}. Here, the reversal is limited to a small-size
system. It is found that the peak of the polarization moves towards
stronger magnetic fields for increasing the width of the central
ZnSe layer, while shifts towards an opposite direction for
increasing the width of paramagnetic layer. A strong suppression of
the spin up component of the current density is present at high
magnetic fields. It is expected that such a reversal of the
polarization could act as a mechanism for a possible selective spin
filter device.

\end{document}